# A real-time beam current density meter

Junliang Liu, Deyang Yu,[a)] Fangfang Ruan, Yingli Xue, and Wei Wang

*Institute of Modern Physics, Chinese Academy of Sciences, Lanzhou 730000, China*

We have developed a real-time beam current density meter for charged particle beams. It measures the mean current density by collimating a uniform and large diameter primary beam. The suppression of the secondary electrons and the deflection of the beam were simulated, and it was tested with a 105 keV $Ar^{7+}$ ion beam.

Informations about current and current density of charged particle beams are very important in many experiments, such as ion sputtering, beam guiding, and irradiation studies.[1-3] The interceptive Faraday-cup or Faraday-cup array, therefore, are widely employed to obtain the absolute current and profile of beams.[4,5] However, non-interceptive real-time monitoring is highly desirable in experimental situations where physical effects depend on the local current density, such as e.g. in beam guiding through capillaries.[2] Although the residual gas monitor technique can provide non-interceptive beam monitoring by measuring the ionized residual gas, it is vacuum dependent and requires careful calibration.[6,7] Moreover, when weak beams are employed, this technique features large experimental uncertainties. We have developed a novel real-time beam current density meter, which can be used for charged particle beams with low, uniform beam densities, with large diameters (≥5mm), and with total currents ranging from pA to μA.

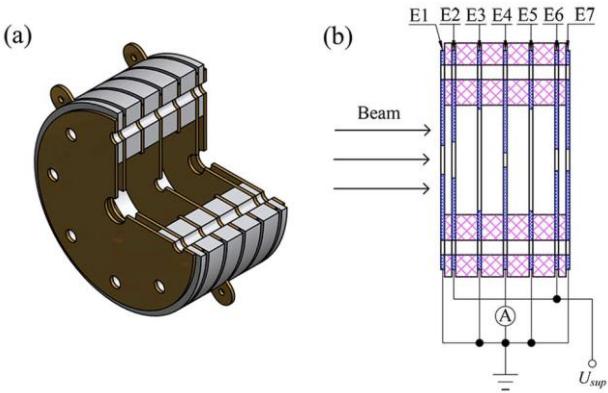

FIG.1. (Color online). Sketch of the real-time beam current density meter. (a) cutaway view; (b) schematic view.

The sketch of the beam current density meter is shown in Fig.1. It is built of 7 coaxially aligned brass electrodes with differently sized central apertures, and which are mutually insulated by using Teflon spacers. The beam enters the current density meter passing through electrode E1. It should be noted that the beam cross sections should be larger than the aperture in E1 and the beam profile should be rather uniform. The actual measurement takes place at the electrode E4, where the outer part of the beam is further cut. Secondary electrons emitted form E4 are suppressed by a negative voltage applied to the suppressor electrodes E2 and E6. The electrodes E1 and E7 are grounded to shield the electrostatic field. The interception electrodes E3 and E5 are grounded too, in order to avoid leakage currents from E2 and E6, respectively, to the measurement electrode E4. The design of the device has been chosen according to the following requirements: First, only E1 and E4 should directly cut the beam, where the size of the aperture in E4 should adapt to the beam size required for the actual experiment. Second, the meter should be as short as possible in the beam propagation direction in order to avoid the direct interaction of the beam with the other electrodes due to the finite beam divergence. And third, secondary electrons have to be suppressed efficiently. Therefore, the apertures of the interception electrodes E3 and E5 are preferably big, while the apertures of the suppressor electrodes E2 and E6 should be only slightly larger than the apertures of E1 and E4, respectively. Furthermore, the distance between the suppressor electrodes and the measurement electrode should be larger than diameters of their apertures, in order to produce efficient suppressing fields. In present case, the aperture diameters of E1-E7 are 4 mm, 5 mm, 14 mm, 2 mm, 14 mm, 3 mm and 3 mm, respectively. The thicknesses of the isolators between E1-E7 are 1 mm, 3 mm, 3 mm, 3 mm, 3 mm and 1 mm, respectively. The thickness of the electrodes is 0.5 mm, which is sufficient to stop ions with MeV energies (the range of e.g. 10 MeV argon ions in copper is 2.44 μm, simulated by the SRIM program[8]). For a rather uniform beam profile inside the aperture of E1, the mean beam density is given by $J = I_m/(A_1 - A_4)$, where $I_m$ is the beam current measured with a conventional picoamperemeter on electrode E4, and $A_1$ and $A_4$ are the areas of the apertures of E1 and E4, respectively. We note that the sensitivity of the picoamperemeter defines the minimum beam current density required for a reliable operation.

[a)]Author to whom correspondence should be addressed. Electronic mail: d.yu@impcas.ac.cn.

Due to heat deposition on E1 and E4, in turn, the beam current densities should also not exceed a critical value.

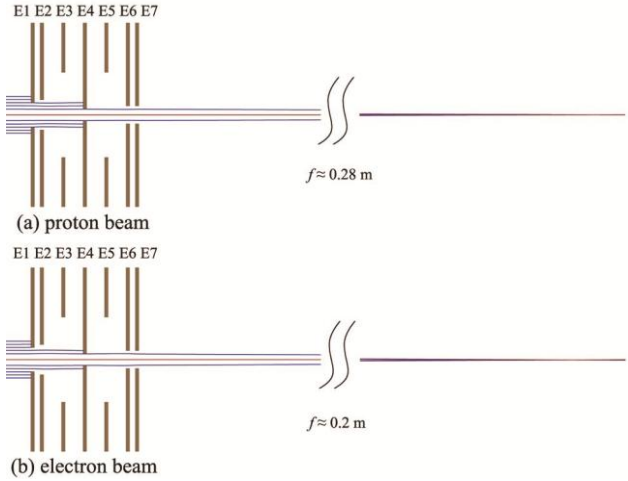

FIG.2. (Color online). Simulated trajectories of (a) 1000eV protons and (b) 1000eV electrons through the beam current density meter. The suppressing voltage on electrodes E2 and E6 is -200 V. The focal length $f$ is defined by the distance from the center of the current density meter to the focal point.

Due to the suppressing voltage, the beam current density meter also acts as an electrostatic lens, and could deflect the beam. While the analytic calculation of the trajectories and the focal length are complicated,[9] numerical methods and algorithms available today provide accurate and simple solutions.[10,11] With the SIMION 8.0 program,[12] we simulated the trajectories of proton and electron beams, as well as the suppression effect on secondary electrons. In the simulations, relatively dense grids of 10 grid units/mm are used which results in a reasonable accuracy of 0.3%.[13]

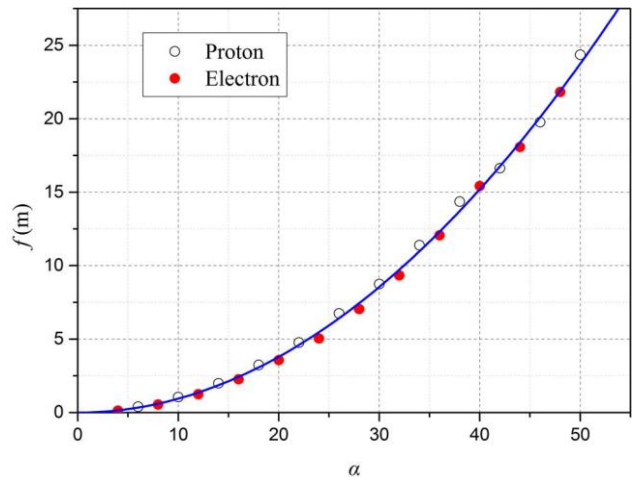

FIG. 3. (Color online). Focal length of the present design as a function of the ratio α (see text). Open and closed circles represent proton and electron beams, respectively. The solid line shows the parabolic increase of the focal length.

In Fig.2, the simulated trajectories of 1000 eV protons and electrons are shown for a suppression voltage of -200 V. Compared to the experiment, a much lower beam energy is chosen in the graph in order to illustrate the focusing effect of the device. We further performed simulations for three suppressing voltages -50 V, -100 V, and -200 V and the ratio $\alpha = |U_{acc}/U_{sup}|$ covering $4 \leq \alpha \leq 50$ ($U_{acc}$ is the beam acceleration voltage and $U_{sup}$ is the suppressing voltage, respectively), as shown in Fig.3. The focal length $f$ roughly depends on the ratio $\alpha$, rather than on the mass and the charge of the particles. When $\alpha$ is greater than 25, the difference of the focal length between proton and electron beam is smaller than 10%, and both are longer than 5 meter. In our applications, the acceleration voltage of the beam is always not less than 15 kV (i.e., $\alpha \geq 75$ when the suppressing voltage is -200 V), and the target is placed about 0.1 meter behind the current density meter. As a result, the deflection of the beam is negligible.

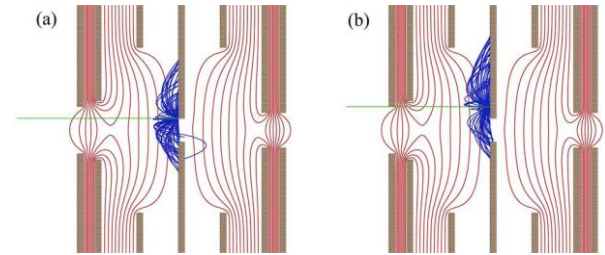

FIG. 4. (Color online). Simulations of secondary electron trajectories (blue lines) in the beam current density meter. The straight lines represent the incident charged particles. The energy of secondary electrons is 50 eV, and the bias voltages on electrodes E2 and E6 is -200 V. The secondary electrons, which are emitted with a random angular distribution in the backward half plane in the graph, were launched from two points on E4: (a) at the inner edge of E4, and (b) at the outer edge of the intersection of the beam with E4. Space charge effects are not considered and expected to be negligibly small.

The kinetic energy distribution of secondary electrons emitted from the measurement electrode E4 depends on the incoming beam species and energy, but electron energies typically do not exceed several ten eV with a distribution maximizing at less than 30 eV.[11] To simulate the suppression of secondary electrons produced on E4, we simulated the trajectories for two groups of electrons that are launched from two extreme points at E4: One position is the edge of the aperture of E4 (Fig.4(a)) and another is the outer edge of the intersection of the beam with E4 (Fig.4(b)). In each group, 100 electrons are emitted with 50 eV with a random angular distribution in the backward half plane in the graph. The bias voltage of electrodes E2 and E6 is set to -200 V. As can be seen in Fig.4, all the electrons are deflected back to the measurement electrode E4, indicating an efficient suppression effect.

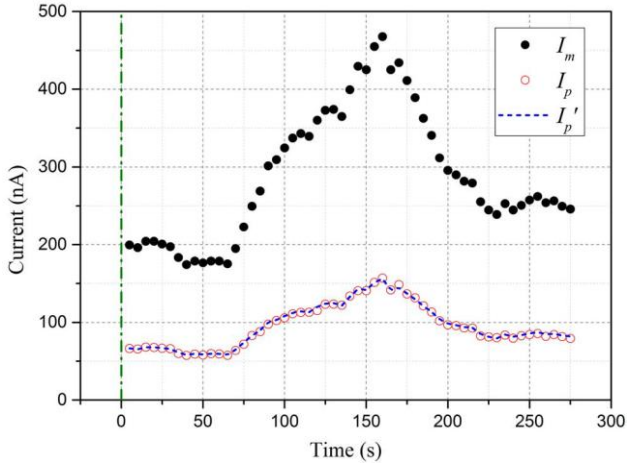

FIG. 5. (Color online). Test measurement of the real-time beam current. The solid circles represent the measured current $I_m$ on the measurement electrode E4 of the density meter, the open circles correspond to the direct measurement of the passing current $I_p$ by a Faraday-cup behind the current density meter, and the dashed line shows the deduced passing current $I_p'$.

A test measurement was carried out at the 320 kV platform for multi-discipline research with highly charged ions[14] at the Institute of Modern Physics, with a beam of 105 keV $Ar^{7+}$ ions. The density meter was placed in a vacuum chamber of $10^{-8}$ mbar. For the experiment the same suppressing voltage of -200 V was used as in the simulations. In front of the chamber a rather uniform beam spot with a diameter about 10 mm diameter was observed on a fluorescent target before the measurement. The real-time measurement current $I_m$ on electrode E4 and the passing current $I_p$ (measured by a Faraday-cup) were recorded simultaneously. $I_m$, $I_p$ and the deduced passing current $I_p' = I_m A_4 / (A_1 - A_4)$ are plotted in Fig.5. During the measurement the beam spot was slightly steered with a dipole magnet. No effect on the passing current was observed, which supports our presumption of a roughly uniform beam current density in the measurement. By adjusting the parameters of the ion source, the beam intensity was varied during the experiment. As can be seen from Fig.5, an excellent agreement between the measured current $I_p$ and the deduced current $I_p'$ has been achieved.

In conclusion, we present a real-time beam current density meter, which operates by collimating a uniform and large diameter primary beam. The focusing of the beam and the suppression of secondary electrons were simulated. It has been tested with a 105 keV $Ar^{7+}$ ion beam. So far, it has been exposed to the beam for more than 200 hours in total at different beam intensities and energies.

We thank the staff of the 320 kV platform for multi-discipline research with highly charged ions of Institute of Modern Physics for their technical support during the experiment. We are grateful to D. Fischer for carefully reading the manuscript. This work was supported by the National Natural Science Foundation of China under Grant No. 11275240 and the Major State Basic Research Development Program of China under Grant No. 2010 CB832901.